\documentclass[12pt]{article}
\usepackage[T1]{fontenc}
\usepackage[utf8]{inputenc}
\usepackage{lmodern}
\usepackage{microtype}
\usepackage[a4paper,margin=0.82in]{geometry}
\usepackage{amsmath,amssymb,bm,mathtools}
\usepackage{graphicx}
\usepackage{booktabs,tabularx,array,multirow}
\usepackage{enumitem}
\usepackage{caption}
\usepackage{subcaption}
\usepackage{siunitx}
\usepackage{xcolor}
\usepackage{hyperref}
\usepackage[numbers,super,sort&compress]{natbib}
\usepackage{authblk}
\usepackage{float}
\usepackage{placeins}
\usepackage{url}
\usepackage{bm}
\usepackage{authblk}
\usepackage{abstract}

\hypersetup{colorlinks=true,citecolor=blue!55!black,linkcolor=blue!55!black,urlcolor=blue!55!black}
\setlist{nosep}
\renewcommand{\arraystretch}{1.15}

\title{\textbf{Symmetry-driven in-situ physical adjoint computing beyond reciprocity}}
\author[1]{Cunshen Li}
\author[1]{Rong Li}
\author[2]{Jiongning Che}
\author[1]{Cheng-Zhen Wang\thanks{Correspondence: wangcz@lzu.edu.cn}}
\author[1]{Liang Huang\thanks{Correspondence: huangl@lzu.edu.cn}}

\affil[1]{Lanzhou Center for Theoretical Physics, Key Laboratory of Quantum Theory and Applications of MoE, Key Laboratory of Theoretical Physics of Gansu Province, Gansu Provincial Research Center for Basic Disciplines of Quantum Physics, and School of Physical Science and Technology, Lanzhou University, Lanzhou 730000, Gansu, China}
\affil[2]{Yangtze Delta Region Institute, University of Electronic Science and Technology of China, Huzhou 313000, Zhejiang, China}
\date{}

\begin{document}
\maketitle
\vspace{-3em}

\begin{abstract}
Nonreciprocity enriches wave control and information processing through directional transport, yet challenges conventional in-situ physical adjoint optimization, which relies on reciprocity to acquire adjoint responses. Here we demonstrate symmetry-driven in-situ physical adjoint computing (Sym-iPAC) in nonreciprocal systems, which enables access to adjoint responses on the original hardware through three classes of transpose symmetry: local geometric permutations, complex gauge reweighting, and latent symmetry in effective port dynamics revealed by Schur reduction. Experiments with programmable nonreciprocal microwave networks validate these transpose relations and yield adjoint gradients, enabling closed-loop optimization. These results establish experimentally accessible transpose symmetry as a route to in-situ physical adjoint computing beyond reciprocity.
\end{abstract}


\section{Introduction}

Nonreciprocity breaks the interchange symmetry between excitation and observation channels, enabling directional wave transport and functionalities unavailable in reciprocal systems\cite{potton2004,svendsen2013,caloz2018,asadchy2020}. In linear time-invariant systems with consistently normalized propagating ports, this asymmetry is manifested by a nonsymmetric scattering matrix, $S\neq S^{\mathrm T}$\cite{potton2004,svendsen2013,caloz2018,asadchy2020}. Nonreciprocal interactions underpin isolation and circulation in microwave and optical platforms\cite{adam2002,bi2011,jalas2013,kord2020}, and have enabled unidirectional topological transport, acoustic circulation and high-isolation microwave devices\cite{fleury2014,wang2009,wang2025isolator}. More recently, nonreciprocity has been exploited for pulse routing and independently controlled bidirectional analogue computing\cite{yang2020lidar,li2025nonreciprocal}. As wave systems become more complex, optimizing their directional responses becomes an important challenge.

Numerical inverse design combines physical models with adjoint sensitivities to optimize complex wave devices\cite{lalau2013,molesky2018,piggott2015,schul2026}. However, fabrication deviations and incompletely characterized scattering interactions can separate model predictions from the response of the realized device\cite{nikkhah2024,sol2024}. Measured responses also enable wave control through experimentally calibrated models of programmable complex media~\cite{sol2024} and feedback-based wavefront shaping~\cite{vellekoop2007}. When gradients are estimated experimentally, coordinate-wise finite
differences require perturbation measurements whose number generally grows with the number of tunable parameters, whereas simultaneous perturbation methods reduce evaluations at the cost of stochastic gradient estimates\cite{spall1992}. Adjoint methods offer a different scaling: for a fixed number of excitation and objective states, sensitivities to all design parameters can, in principle, be obtained from a small number of global field solutions, independent of parameter count at the level of global solves\cite{lalau2013,molesky2018}. In-situ adjoint computing aims to combine this scaling advantage with measured responses from the actual device. Its implementation in nonreciprocal systems, however, requires physical access to the adjoint dynamics on the original hardware.

Related approaches to physical training have explored
both backpropagation-based and backpropagation-free
methods~\cite{Wright2022Deep,Momeni2023Backpropagation,Momeni2025Training}. Physical backpropagation has progressed from proposals for in-situ photonic gradient measurement\cite{hughes2018insitu,zhou2020diffractive,thakkar2026circuits} to experimental demonstrations in integrated photonic meshes\cite{pai2023insitu}. Fully forward mode training exploits spatial symmetry together with Lorentz reciprocity\cite{xue2024ffm}, while in-situ physical adjoint computing has enabled adjoint assisted optimization in reciprocal multiple-scattering microwave environment\cite{guillamon2025ipac}. Recent optical experiments have demonstrated on-chip backpropagation and synthetic time-reversal holography\cite{ashtiani2026onchip,zhou2026inspire}. Related studies have explored in-situ time-domain circuits\cite{kwon2026time}, nonlinear $\mathcal{PT}$-symmetric systems\cite{li2026pt} and stochastic-adjoint gradient estimation~\cite{Tuxbury2026Physical}. A Faraday-adjoint method has been theoretically formulated and numerically validated for inverse design in magneto-optical media, relying on computational responses
for reversed magnetic bias rather than in-situ adjoint propagation on the operating device\cite{cisowski2024faraday}. A contemporaneous theoretical study has provided a unified adjoint framework for in-situ physical backpropagation, including extensions to a class of nonreciprocal systems~\cite{bosch2026unifying}. Moreover, Guillamon et al. developed a
phase-engineered adjoint protocol and provided the first
experimental demonstration of in-situ adjoint optimization
in nonreciprocal microwave systems, with nonreciprocity
arising from a localized nonlinear
defect~\cite{guillamon2026nonlinear}.

Reciprocity ensures that forward and adjoint wave propagation
are governed by the same operator. In nonreciprocal systems,
this direct correspondence is generally lost, and even when
a more general transpose mapping exists, it does not by
itself guarantee physical access to the adjoint response.
A practical mapping must remain fixed throughout parameter tuning,
connect experimentally accessible excitations and measurements,
and enable the recovery of local sensitivities required for
gradient evaluation. Importantly, such a mapping need not rely
on global geometric symmetry: it may act only on selected
degrees of freedom or emerge in the effective port dynamics
after eliminating inaccessible internal states.
The central challenge is therefore to realize these structured
transpose mappings on the original nonreciprocal hardware
and reconstruct reliable adjoint gradients from measured
responses for closed-loop optimization.

Here we demonstrate symmetry-driven in-situ physical adjoint computing (Sym-iPAC) in linear nonreciprocal microwave
systems, without constructing a separate transposed device. We derive criteria for parameter-independent transpose
mappings through geometric permutations and complex gauge transformations, and show how Schur reduction reveals latent transpose symmetry in accessible-port dynamics by eliminating inaccessible internal degrees of freedom. These mappings can accommodate independently tunable reciprocal subsystems without requiring
global geometric symmetry. We test these mechanisms in four programmable
coaxial microwave graphs containing nonreciprocal components. Measured scattering matrices test the predicted port-level relations, while digital superposition of sequential single-port measurements reconstructs the transformed adjoint fields. Calibrated system local relations provide the operator derivatives needed for gradient evaluation, without relying on full system simulations. We validate the resulting stabilized in-situ adjoint gradients against independent finite-difference measurements and use them to drive closed-loop optimization on the original hardware. The optimized directional transmission contrasts reach final values comparable to those obtained with live finite-difference control. These results establish experimentally accessible transpose symmetry as a route to in-situ adjoint-based optimization beyond Lorentz reciprocity.

\section{Symmetry-driven in-situ adjoint beyond reciprocity}
\label{sec:adjoint_access}

\begin{figure}[t]
\centering
\includegraphics[width=0.95\textwidth]{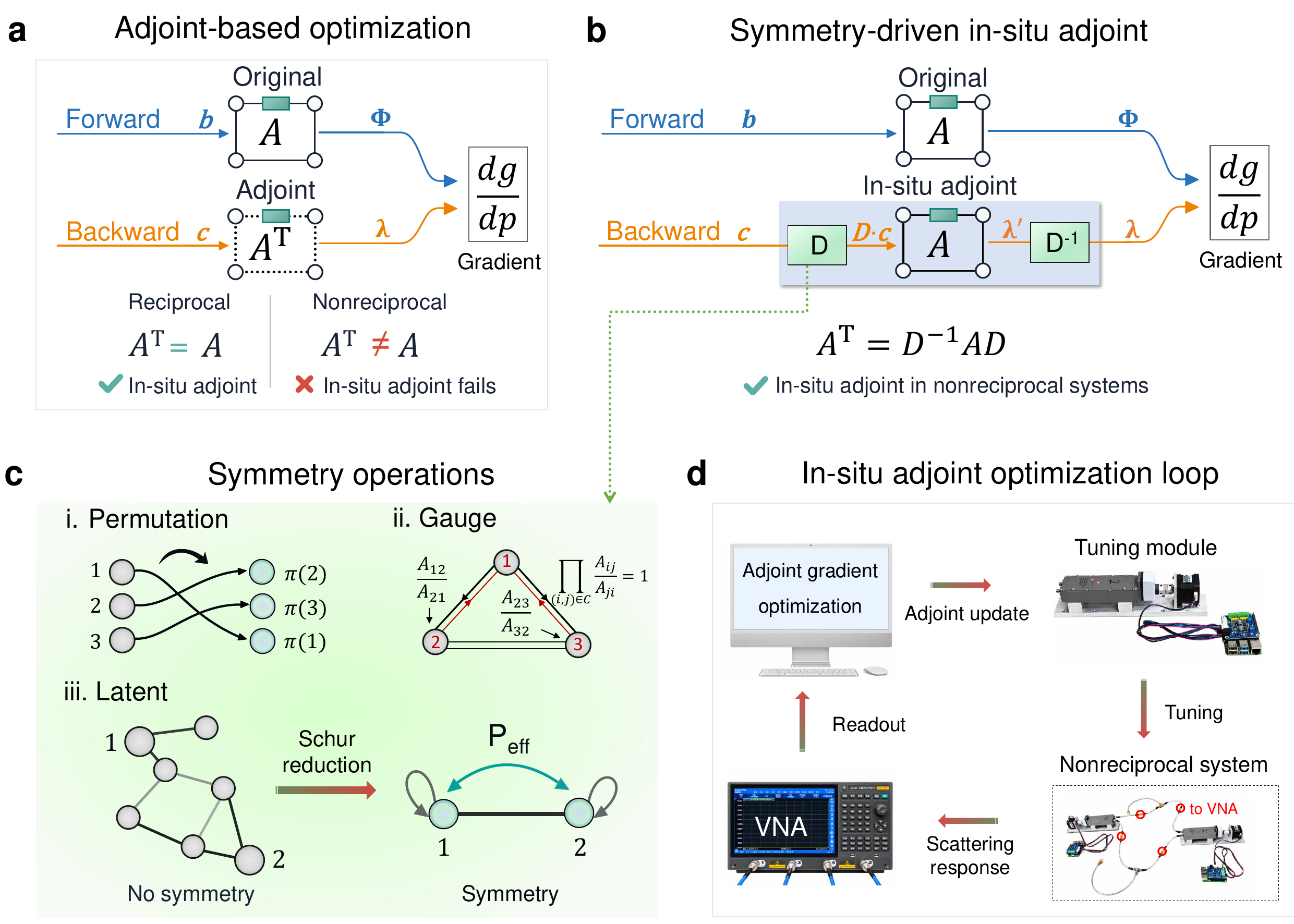}
\caption{\textbf{Symmetry-driven in-situ adjoint computing
in nonreciprocal wave systems.}
\textbf{a.} The forward and adjoint fields satisfy
$A\bf\Phi=b$ and $A^{\mathrm T}\boldsymbol{\lambda}=\boldsymbol{c}$, respectively.
In reciprocal systems, $A=A^{\mathrm T}$ allows
direct adjoint access, whereas in nonreciprocal
networks, $A\neq A^{\mathrm T}$ generally prevents
adjoint recovery by simply exchanging excitation
and observation channels.
\textbf{b.} A structured mapping
$DA^{\mathrm T}D^{-1}=A$ enables the adjoint response
to be reconstructed on the original hardware through
transformed excitation and field recovery, without
reversing the magnetic bias.
\textbf{c.} Three mechanisms provide accessible
transpose mappings: geometric permutation,
complex gauge reweighting and latent
symmetry revealed in the effective port dynamics
by Schur-complement reduction.
\textbf{d.} A four-port vector network analyzer
sequentially excites the accessible ports and
records complex scattering responses. A control
computer digitally reconstructs the adjoint response
and computes update directions, while motorized
phase shifters adjust branch lengths for closed-loop
optimization.}
\label{fig:concept}
\end{figure}

Gradient-based optimization of a wave system requires
both forward and adjoint (backward) responses
(Fig.~\ref{fig:concept}a). At a fixed frequency,
these fields satisfy
\begin{equation}
A(\mathbf{p})\boldsymbol{\Phi}=\mathbf{b},
\qquad
A^{\mathrm{T}}(\mathbf{p})\boldsymbol{\lambda}
=\mathbf{c},
\label{eq:forward_adjoint}
\end{equation}
where $A(\mathbf{p})$ is the parameter-dependent wave
operator, $\boldsymbol{\Phi}$ and $\boldsymbol{\lambda}$
are the forward and adjoint fields, and $\mathbf{b}$
and $\mathbf{c}$ denote the forward and
objective-dependent adjoint (backward) sources, respectively.
The superscript ${\mathrm T}$ denotes ordinary
transpose rather than conjugate transpose.
For fixed excitations, parameter gradients follow
from local contractions of the forward and adjoint
fields with derivatives of the wave operator
(Methods). In reciprocal systems, the adjoint
response can be accessed by exchanging excitation
and observation channels. Nonreciprocity generally
breaks this correspondence because
$A^{\mathrm T}\neq A$, preventing direct in-situ adjoint
access through conventional port interchange.

Rather than physically reproducing the transposed
network, we seek a fixed, invertible transformation
$D$ satisfying
\begin{equation}
D A^{\mathrm{T}}(\mathbf{p})D^{-1}
=A(\mathbf{p}),
\qquad \forall\,\mathbf{p}\in\mathcal{P},
\label{eq:transpose_map}
\end{equation}
where $\mathcal{P}$ denotes the admissible
parameter space. Defining the transformed adjoint
field $\boldsymbol{\lambda}'=D\boldsymbol{\lambda}$
gives
$A(\mathbf{p})\boldsymbol{\lambda}'=D\mathbf{c}$,
with
$\boldsymbol{\lambda}=D^{-1}\boldsymbol{\lambda}'$.
Thus, the original physical network can supply the
transformed adjoint response without constructing a separate transposed
device (Fig.~\ref{fig:concept}b). 
Crucially, this algebraic equivalence does not
by itself guarantee physical accessibility:
the transformed excitation must be realizable
through accessible channels, the required field
information must be recoverable, and the same
transformation must remain valid throughout
parameter tuning. Accessible transpose mappings can arise from geometric
permutations or complex gauge reweighting, while
Schur-complement reduction can reveal latent transpose
symmetry in the effective dynamics by eliminating
internal degrees of freedom while retaining the
accessible ports (Fig.~\ref{fig:concept}c).

\subsection{Permutation transpose symmetry}

The simplest transpose mapping is realized by a fixed
permutation matrix $P$ that permutes the system's
degrees of freedom. Transposition exchanges the coupling elements $A_{ij}$
and $A_{ji}$, while the permutation maps the transposed
operator back onto the original system:
\begin{equation}
P A^{\mathrm T}(\mathbf p)P^{-1}=A(\mathbf p),
\qquad \forall\,\mathbf p\in\mathcal P.
\end{equation}
Although this condition applies to the complete
operator, it does not require global geometric
symmetry. For a three-terminal nonreciprocal
component, the permutation can exchange the
connections to two terminals while leaving the
third terminal and its attached reciprocal
subsystem unchanged. The latter can have
arbitrary reciprocal couplings and independently
tunable parameters, provided its interfaces
remain compatible with the permutation.
The mapping must remain remain fixed throughout the admissible tuning range,
rather than hold only at a single parameter setting.

When the permutation maps the required adjoint source
and readout onto experimentally accessible ports, the
transformed adjoint response can be obtained by exciting
the corresponding ports on the original network and
recovering the adjoint field through the inverse
permutation. No physical reversal of the nonreciprocal component (i.e., circulator) is required; $P$ represents a correspondence between system's degrees of freedom,
not a physical transformation of the device. Mirror reflection (Fig.~\ref{fig:concept}c(i), Fig.~\ref{fig:symmetry_mappings}a,d) and rotation (Extended Data Fig.~1) provide examples of
such mappings.
The general criterion and constructive procedure
for identifying the required permutation are provided
in the Supplementary Information (SI)~\cite{symipac2026SI}.

\subsection{Gauge transpose symmetry}

A complementary route to transpose symmetry is gauge
reweighting, in which a diagonal complex transformation
compensates for directional coupling asymmetries without
relabeling the system's degrees of freedom(Fig.~\ref{fig:concept}c(ii), Fig.~\ref{fig:symmetry_mappings}b,e). We consider
$D=\Lambda=\operatorname{diag}(d_1,\ldots,d_N)$, with
$d_i\neq0$. The transpose relation in Eq.~(2) requires
\begin{equation}
r_{ij}(\mathbf p)\equiv
\frac{A_{ij}(\mathbf p)}{A_{ji}(\mathbf p)}
=\frac{d_i}{d_j},
\qquad \forall\,\mathbf p\in\mathcal P,
\label{eq:gauge_local_ratio}
\end{equation}
for every pair of degrees of freedom with nonzero
coupling in both directions. A diagonal transformation
cannot change the coupling support, so the two
directions must have matching support:
$A_{ij}(\mathbf p)=0$ if and only if
$A_{ji}(\mathbf p)=0$. Diagonal entries impose no
additional constraint because they are unchanged
by transposition and diagonal similarity. Importantly, reciprocal couplings between degrees
of freedom with identical gauge weights can be
freely designed and tuned without changing the
transpose mapping. Thus, the required symmetry
does not impose geometric constraints on these
reciprocal substructures, allowing them to remain
structurally complex and independently tunable.

Constructing a fixed mapping requires two further
conditions. First, each coupling ratio must be
independent of the tunable parameters, so that
$r_{ij}(\mathbf p)=r_{ij}^{(0)}$ throughout the
admissible parameter range wherever the coupling
is nonzero. To construct the weights, we choose
an arbitrary nonzero reference value for one degree
of freedom in each connected component and propagate
the ratios through the coupled degrees of freedom.
For a coupling pair $(i,j)$, if $d_i$ is known, we
set $d_j=d_i/r_{ij}^{(0)}$. The resulting weights
are defined up to a common nonzero scale within each
connected component.

Second, the propagated weights must be independent
of the path used to reach each degree of freedom.
This requires the product of coupling ratios around
every closed coupling loop $C$ to equal unity:
\begin{equation}
\prod_{(i,j)\in C}r_{ij}^{(0)}=1.
\label{eq:gauge_cycle_consistency}
\end{equation}
If this condition fails, no diagonal transformation
satisfying Eq.~(2) exists. Conversely, matched coupling
support, parameter-independent ratios and consistency
around all closed loops provide a constructive
criterion for a fixed diagonal mapping. Cycle
consistency at each parameter setting alone is
insufficient: it may permit a different transformation
at each setting without yielding a single fixed
mapping over $\mathcal P$.

Because the weights are complex and need not have
unit modulus, gauge reweighting can compensate both
amplitude and phase asymmetries. When geometric
relabeling is also required, the two operations can
be combined as $D=P\Lambda$, provided the complete
operator satisfies Eq.~(2). The general constructive
criterion and its proof are given in Methods and
SI~\cite{symipac2026SI}.

\subsection{Schur reduction and latent transpose symmetry}

In many physical systems, only a subset of the degrees
of freedom is accessible to excitation and measurement.
Partitioning the system into retained states $r$ and
inaccessible internal states $i$, with no independent
source applied to the latter, allows the internal
amplitudes to be eliminated through the Schur complement:
\begin{equation}
A_{\mathrm{eff}}(\mathbf p)
=A_{rr}(\mathbf p)
-A_{ri}(\mathbf p)A_{ii}^{-1}(\mathbf p)A_{ir}(\mathbf p),
\qquad
A_{\mathrm{eff}}\boldsymbol{\Phi}_r=\mathbf b_r.
\label{eq:main_schur_operator}
\end{equation}
The effective operator retains the feedback mediated
by the eliminated states; this is a mathematical
reduction, not the physical removal of the internal
subsystem.

Consider a fixed full space transpose mapping
$DA^{\mathrm T}D^{-1}=A$, with $D$ partitioned
according to the retained and internal states.
If $D$ preserves this partition, so that
$D=\operatorname{diag}(D_r,D_i)$, the mapping
is inherited by the reduced operator:
$D_rA_{\mathrm{eff}}^{\mathrm T}D_r^{-1}
=A_{\mathrm{eff}}$.
However, if $D$ mixes retained and internal states,
the reduced relation generally takes the two-sided
form
$A_{\mathrm{eff}}D_{\mathrm{fld}}
=D_{\mathrm{src}}A_{\mathrm{eff}}^{\mathrm T}$,
where $D_{\mathrm{src}}$ and $D_{\mathrm{fld}}$
describe the effective source transformation and
field transformation, respectively (Methods).
These transformations need not coincide and may
depend on $\mathbf p$ even when the full-space
$D$ is fixed. Thus, the retained block $D_{rr}$
alone does not generally define a valid,
parameter-independent transpose mapping.

More generally, the reduced operator can possess
its own fixed transpose symmetry even when no
corresponding geometric symmetry is apparent in
the full system. We call this a latent transpose
symmetry\cite{kempton2020isospectral,rontgen2021latent,
rontgen2023equireflectionality}(Fig.~\ref{fig:concept}c(iii), Fig.~\ref{fig:symmetry_mappings}c,f), defined by an invertible transformation
acting only on the retained states:
\begin{equation}
D_{\mathrm{eff}}A_{\mathrm{eff}}^{\mathrm T}(\mathbf p)
D_{\mathrm{eff}}^{-1}
=A_{\mathrm{eff}}(\mathbf p),
\qquad \forall\,\mathbf p\in\mathcal P.
\label{eq:main_latent_transpose}
\end{equation}
This condition must hold throughout the admissible
parameter range. In our latent-symmetry system,
reduction to the four accessible ports reveals
$D_{\mathrm{eff}}=P_{12}$, which exchanges ports 1
and 2 while leaving ports 3 and 4 unchanged over
the measured $\Delta L_{78}$ scan at 8.3700 GHz.
The relation is verified across the full effective
operator, rather than inferred from selected matrix
elements alone.

A latent port mapping provides access to the
transformed adjoint response, but does not by itself
determine the local fields required for gradients
of internal parameters. These sensitivities require
effective-operator derivatives or calibrated local-field
relations (Methods). In our implementation, the
transformed port response is reconstructed by linear
superposition of sequential single-port measurements,
while calibrated local relations provide the additional
information needed for gradient evaluation. The general
reduction, mixed source-field mappings and their
derivation are detailed in SI~\cite{symipac2026SI}.

\section{From transpose symmetry to in-situ adjoint optimization}
\label{sec:experimental_results}

\subsection{Accessible transpose mappings in nonreciprocal microwave network systems}

\begin{figure}[t]
\centering
\includegraphics[width=0.98\textwidth]{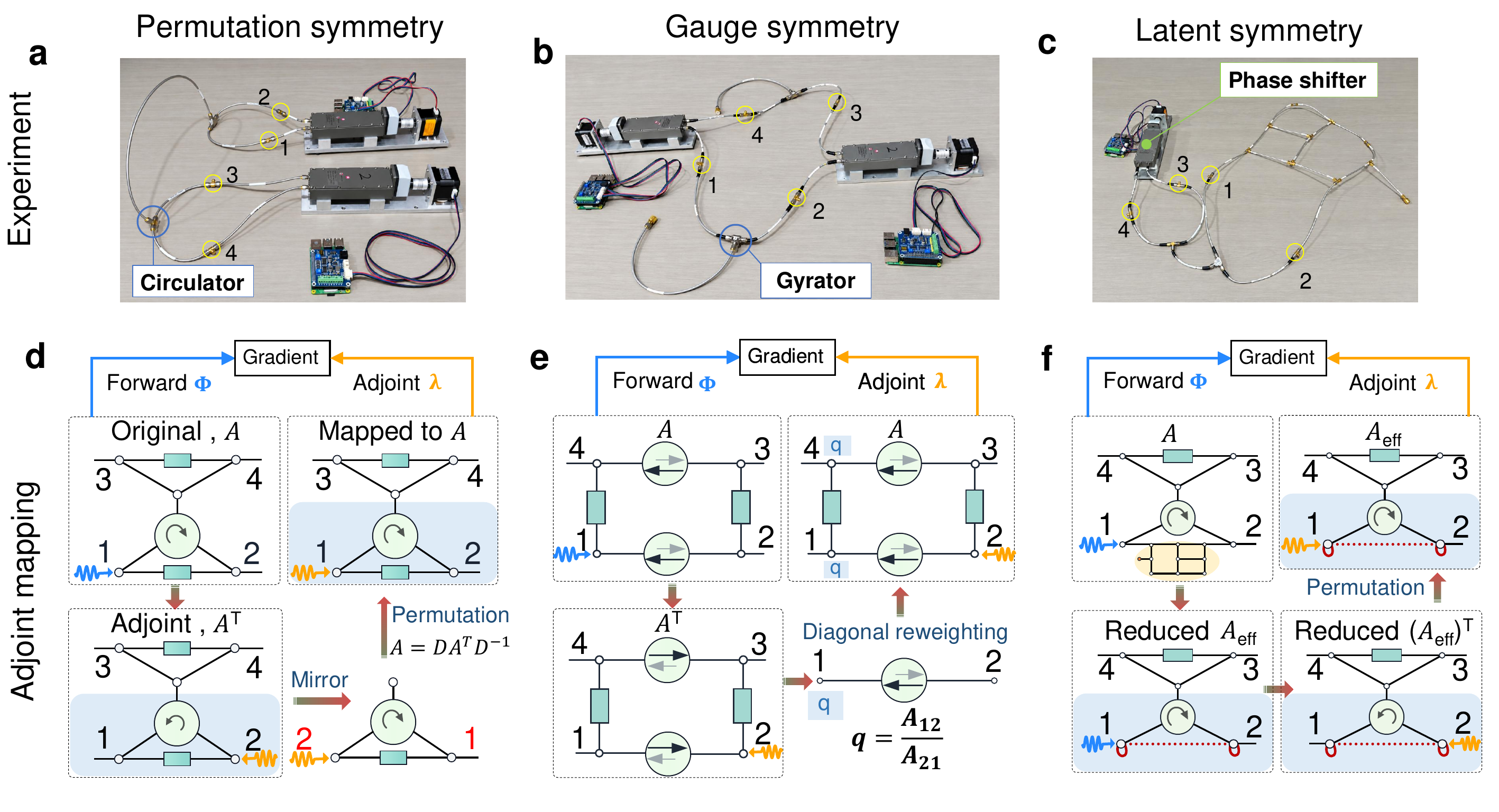}
\caption{\textbf{Accessible transpose mappings in
nonreciprocal microwave networks.}
\textbf{a-c}, Photographs of the permutation symmetry
(\textbf{a}), gauge symmetry (\textbf{b}) and
latent symmetry (\textbf{c}) microwave networks.
Yellow circles indicate accessible ports, green dots
tunable phase shifters and blue circles
nonreciprocal components.
\textbf{d}, Mirror reflection maps the transposed
dynamics onto the original network through a fixed
geometric permutation, inducing the port exchange
$P_{12}$.
\textbf{e}, Complex amplitude-phase reweighting
provides the gauge symmetry mapping
$D_{\mathrm p}=\operatorname{diag}(q,1,1,q)$,
where $q\simeq0.8720+0.0989i$.
\textbf{f}, Schur complement reduction reveals a
latent transpose symmetry in the effective port
dynamics, also described by $P_{12}$, despite the
absence of an apparent geometric correspondence
in the full network.
Blue and orange indicate forward and adjoint
excitation, respectively. The transformed adjoint
responses are reconstructed from sequentially
measured complex port responses on the original
hardware.}
\label{fig:symmetry_mappings}
\end{figure}

We first test whether structured transpose mappings
remain consistent with measured responses in
nonreciprocal hardware throughout parameter tuning.
We use three types programmable microwave network systems (quantum graphs)
assembled from coaxial transmission lines (with four external ports), junctions,
ferrite-based nonreciprocal components (providing nonreciprocity) and motorized
phase shifters. The phase shifters tune selected branch
lengths, while a four-port vector network analyzer (VNA) sequentially
excites the accessible ports to measure the complete
complex scattering matrix (Methods).

Three types of networks realize distinct mechanisms for
accessing the transpose dynamics (Fig.~\ref{fig:symmetry_mappings}). In the
permutation symmetry network (Fig.~\ref{fig:symmetry_mappings}a,d), mirror
reflection maps the transposed network onto the
original graph through a port permutation $P_{12}$,
which exchanges ports 1 and 2 while leaving ports 3
and 4 unchanged. The permutation pairs the connections to two
circulator terminals, while the reciprocal
subsystem attached to the invariant third
terminal need not possess mirror symmetry. In the gauge symmetry network
(Fig.~\ref{fig:symmetry_mappings}b,e), complex amplitude-phase reweighting
provides the port transformation
$D_{\mathrm p}=\operatorname{diag}(q,1,1,q)$,
where $q\simeq0.8720+0.0989i$ is determined from
component-level scattering calibration (Methods).
Because $|q|\neq1$, this mapping requires amplitude
as well as phase reweighting. The two tunable reciprocal bonds connect
degrees of freedom with identical gauge weights, allowing their lengths to vary independently
without altering the fixed transpose mapping.
In the latent symmetry
network (Fig.~\ref{fig:symmetry_mappings}c,f), the complete graph lacks an
apparent geometric symmetry, but Schur-complement
reduction reveals the same $P_{12}$ transpose mapping
in the effective port dynamics. The internal network
remains physically present. A rotationally
symmetric network provides an additional realization
of permutation symmetry (Extended Data Fig.~\ref{fig:ed_rotation}).

We test the predicted mappings using the measured
four-port scattering matrix $S$. Their boundary
signature is
\begin{equation}
S D_{\mathrm p}=D_{\mathrm p}S^{\mathrm T},
\end{equation}
which we quantify by the normalized residual
\begin{equation}
r_D(\mathbf p)=
\frac{\left\|
D_{\mathrm p}S^{\mathrm T}(\mathbf p)
D_{\mathrm p}^{-1}-S(\mathbf p)
\right\|_{\mathrm F}}
{\left\|S(\mathbf p)\right\|_{\mathrm F}}.
\label{eq:experimental_port_residual}
\end{equation}
Crucially, the same $D_{\mathrm p}$ is retained
throughout each parameter scan, rather than refitted
at every configuration. The mean residual is 0.0608
for the permutation symmetry network, 0.0115 for the
gauge symmetry network and 0.0677 for the
latent symmetry network (Extended Data Table~1).
The rotational network has a mean residual of 0.1089
(Extended Data Fig.~\ref{fig:ed_rotation} and
Extended Data Table~1). These finite residuals
quantify the degree to which the predicted transpose
relations are realized by the measured devices;
component mismatch, connection errors and calibration
offsets can contribute to the deviations. The measurements establish that distinct structured
mappings can remain consistent with the accessible port
responses across the tested tuning ranges.

\begin{figure}[t]
\centering
\includegraphics[width=0.98\textwidth]{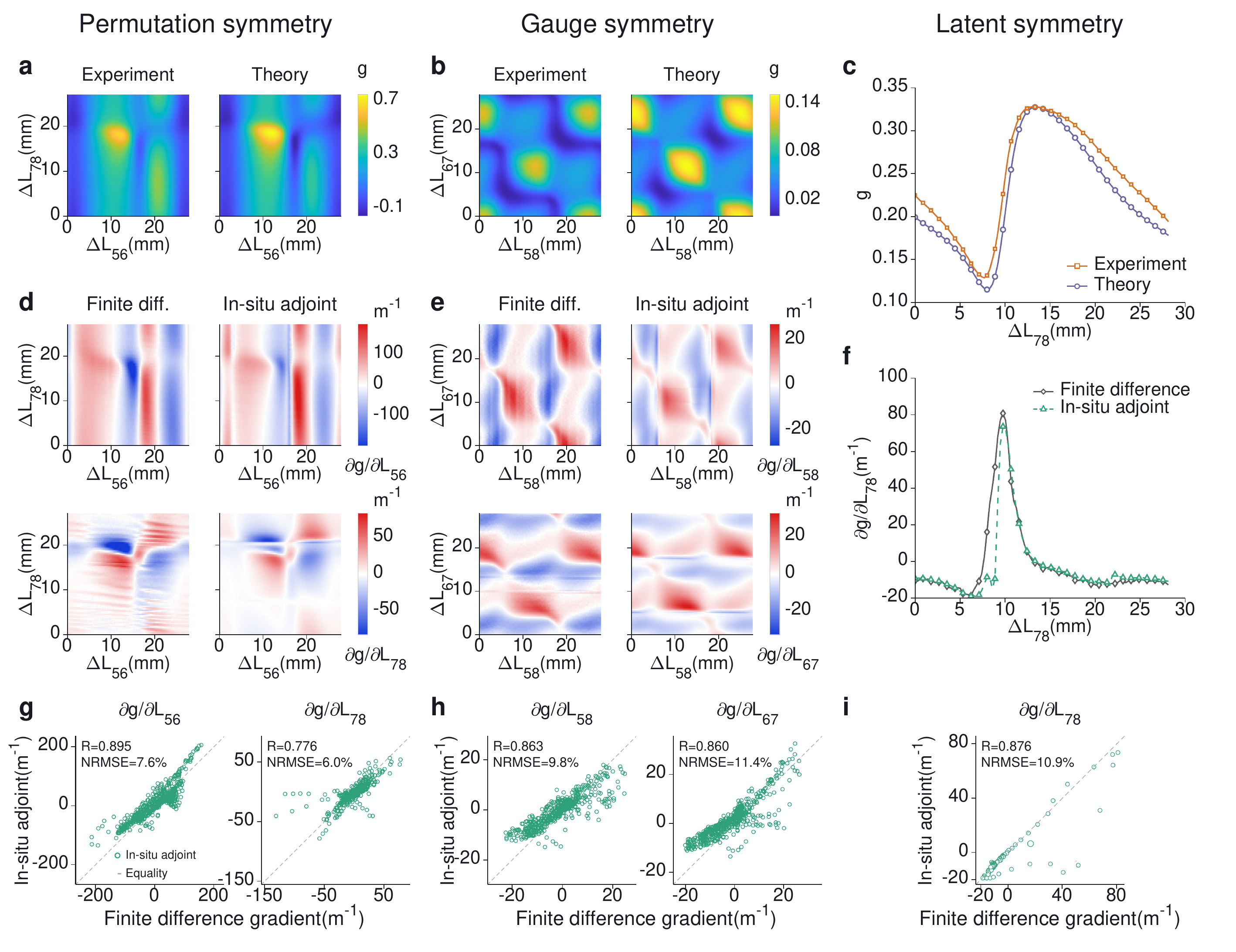}

\caption{\textbf{Experimental validation of in-situ adjoint
gradients in nonreciprocal microwave network systems.}
\textbf{a-c}, Measured and theoretical objectives of the
directional transmission contrast
$g=\sigma(|S_{21}|^2-|S_{12}|^2)$ for the
permutation symmetry (\textbf{a}), gauge symmetry
(\textbf{b}) and latent symmetry (\textbf{c}) networks.
\textbf{d-f}, In-situ adjoint gradients compared with
finite difference estimates for the permutation symmetry
(\textbf{d}), gauge symmetry (\textbf{e}) and latent symmetry
(\textbf{f}) networks. \textbf{g-i}, Pointwise comparisons
of in-situ adjoint gradients with independent finite difference
estimates for the same networks, respectively. Dashed lines
indicate equality. The in-situ adjoint gradients are stabilized
as described in Methods. $R$ denotes the Pearson correlation
coefficient, and NRMSE the range-normalized root-mean-square
error.}
\label{fig3}
\end{figure}

\subsection{Experimental validation of in-situ adjoint gradients}
The port transpose relation is necessary but not sufficient for gradient-based optimization: the
gradient with respect to an internal bond length depends on the local forward and adjoint fields and on the
corresponding derivative of the wave operator. We therefore test whether the symmetry-driven in-situ
adjoint method provides reliable gradient information across the permutation, gauge and latent symmetry
systems. We use the directional transmission contrasts as the optimization objective function, 
\begin{equation}
g_{\sigma}(\mathbf p)=\sigma(|S_{21}|^2-|S_{12}|^2)=
\sigma\left(
|\varphi_{2}^{(1)}(\mathbf p)|^{2}
-|\varphi_{1}^{(2)}(\mathbf p)|^{2}
\right),
\label{eq:experimental_objective}
\end{equation}
where $\mathbf p$ denotes the tunable bond lengths, $\varphi_m^{(n)}$ is the field at accessible
node $m$ under unit excitation from port $n$.
The sign factor is $\sigma=-1$ for the permutation symmetry
network and $\sigma=+1$ for the gauge and
latent symmetry networks, so that the objective has
the desired direction in each configuration.

At each parameter setting, a VNA sequentially excites the accessible ports and
records the complex scattering responses. Linear superposition of this measured port basis, with
coefficients determined by the transformed adjoint source, reconstructs the transformed adjoint field
at the accessible ports. Calibrated local graph relations are used to recover the fields at the endpoints of each tunable bond
and evaluate its corresponding operator derivative, without requiring a complete model-based simulation of the network (Methods). The resulting in-situ adjoint gradient estimates are stabilized before comparison with finite differences and used in optimization (Methods).
Thus, this test evaluates the full reconstruction procedure, from measured port responses to local gradient estimates, rather than inferring gradient accuracy from the transpose relation alone.

The measured and theoretical landscapes of the
directional transmission contrast $g$ capture the
objective function features of the permutation
symmetry (geometric mirror symmetry) network at 8.0276 GHz, scanned over
$(\Delta L_{56},\Delta L_{78})$, and the gauge symmetry
network at 8.6400 GHz, scanned over
$(\Delta L_{58},\Delta L_{67})$. For the latent symmetry
network, $g$ is measured as $\Delta L_{78}$ varies
at 8.3700 GHz (Fig.~\ref{fig3}a-c). Across all three networks,
the in-situ adjoint gradient maps reproduce the main
features of the finite-difference references, including
prominent extrema and sign changes (Fig.~\ref{fig3}d-f).
Comparisons of the in-situ adjoint gradient distributions
with model-derived theoretical gradients are provided
in Extended Data Fig.~2.
These comparisons assess consistency between the
experimentally reconstructed gradients and the calibrated
quantum graph model, complementing the independent
finite-difference benchmark based on measured objective
values.

Pointwise comparisons quantify this agreement (Fig.~\ref{fig3}g-i). For the permutation symmetry network,
the Pearson correlation coefficients are $R=0.895$ and $0.776$ for $L_{56}$ and $L_{78}$, with
NRMSE values of 7.6\% and 6\%, respectively. For the gauge symmetry network, the corresponding
values are $R=0.86282$ and $0.86014$, with NRMSE values of 9.776\% and 11.450\% for $L_{58}$ and
$L_{67}$. For the latent-symmetry network, the comparison gives $R=0.87630$ and an NRMSE of
10.851\%. The additional permutation symmetry (geometric rotational symmetry) network yields $R=0.97330$ and an NRMSE of 6.295\%
(Extended Data Fig.~\ref{fig:ed_rotation} and Extended Data Table~2).

These results show that experimentally accessible transpose mappings can provide useful in-situ
adjoint gradients across distinct symmetry mechanisms. The agreement is significant because
the measured port-level relation alone does not determine the local fields needed for bond length
gradients. The reported gradients include the stabilization described in Methods and are therefore
best interpreted as validated optimization directions, rather than necessarily exact derivatives of the
unmodified objective.

\subsection{Closed-loop optimization on nonreciprocal hardware}

\begin{figure}[t]
\centering
\includegraphics[width=0.98\textwidth]{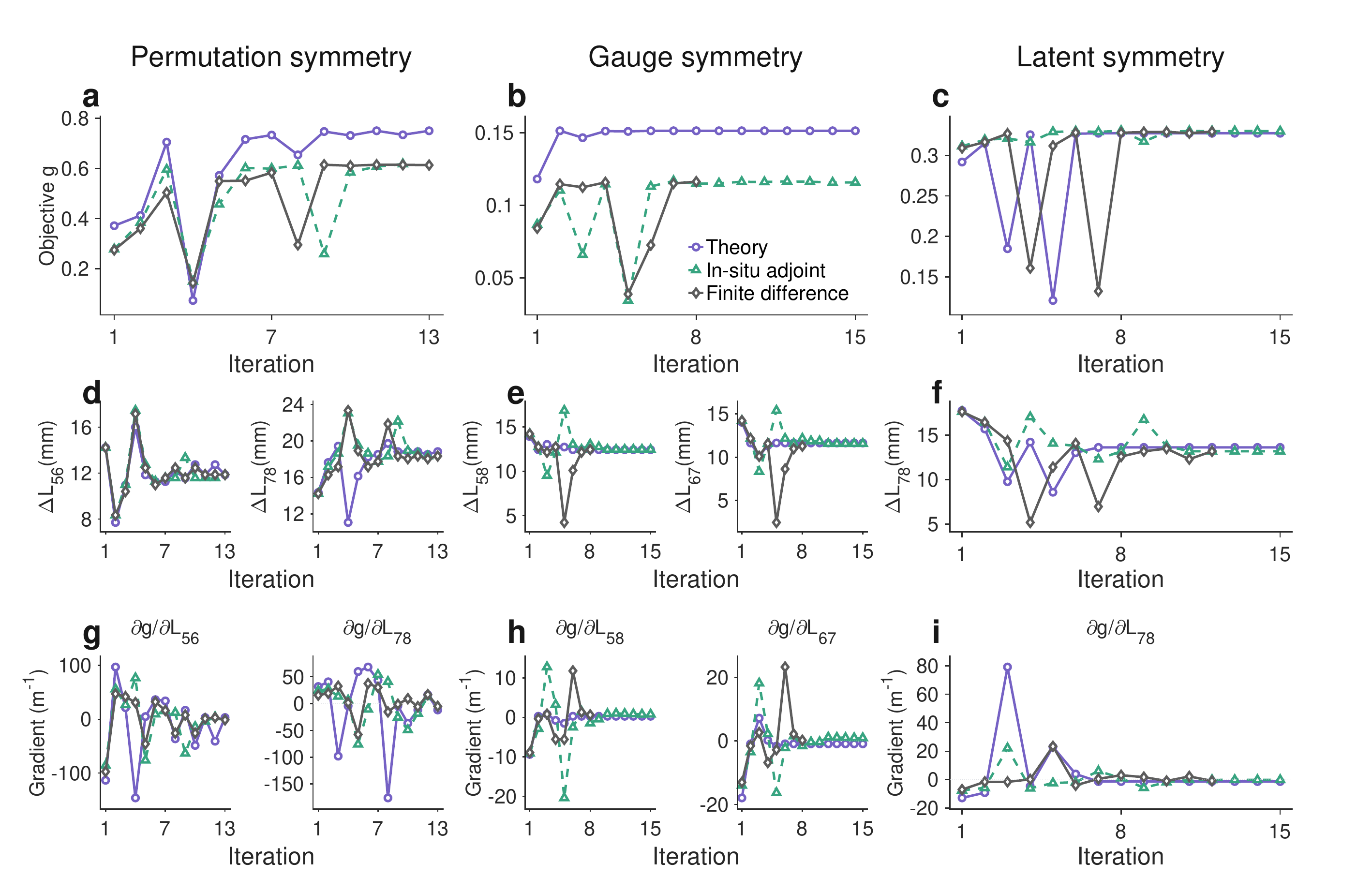}

\caption{\textbf{Closed-loop optimization using in-situ
adjoint gradients in nonreciprocal microwave networks.}
\textbf{a-c}, Evolution of the directional transmission
contrast $g$ during optimization of the permutation
symmetry (\textbf{a}), gauge symmetry (\textbf{b}) and
latent symmetry (\textbf{c}) networks.
\textbf{d-f}, Corresponding evolution of the tunable
bond length adjustments $\Delta L$.
\textbf{g-i}, corresponding in-situ adjoint gradients along
the optimization trajectories. Green triangles and grey diamonds denote
in-situ adjoint and finite difference optimization,
respectively; purple open circles denote an offline
model based reference. Live optimization is performed
on the same nonreciprocal hardware (in-situ). The in-situ adjoint gradients are
stabilized as described in Methods.}
\label{fig4}
\end{figure}

Having validated the in-situ adjoint gradients against independent finite differences, we next test
whether they can guide parameter updates in a live optimization loop. We optimize the directional
transmission contrast on the original permutation, gauge and latent symmetry networks (Fig.~\ref{fig4}). At each iteration, the accessible ports are measured sequentially, and the transformed adjoint field is reconstructed by digital superposition of the measured
complex responses. Calibrated local graph relations supply the field information and operator derivatives required for bond length gradients, without replacing the measured port responses with a full-network forward-adjoint simulation (Methods). The resulting stabilized in-situ adjoint gradients are passed to a moving-asymptotes optimizer\cite{svanberg1987,johnson2021nlopt}, which updates the tunable
bond lengths through programmable phase shifters. The optimization is on the same nonreciprocal device throughout the process.

The in-situ adjoint gradients improve the objective in all three networks, reaching final values comparable
to those obtained by live finite-difference optimization (Fig.~\ref{fig4}a-c). In the permutation symmetry network,
the objective increases from 0.2782 to 0.6174 for in-situ adjoint, compared with 0.6132 using finite differences. In the gauge symmetry
network, it increases from 0.0869 to 0.1151 for in-situ adjoint, compared with 0.1163 using finite differences. In the latent symmetry
network, the recorded final value is 0.3298 starting from 0.3115 along the in-situ adjoint trajectory, compared with 0.3294 using finite differences. The rotational network provides an additional test, reaching 0.6540 from 0.4833 with in-situ adjoint updates and 0.6471 with live finite differences (Extended Data Fig.~\ref{fig:ed_rotation} and Extended Data Table~3). Thus, across distinct mechanisms for accessing the transpose dynamics, the experimentally reconstructed gradients provide useful update directions for optimizing directional transmission on the original nonreciprocal
hardware.

The bond length trajectories (Fig.~\ref{fig4}d-f) and gradient histories (Fig.~\ref{fig4}g-i) show how the parameter updates and their
associated sgradients evolve during optimization. Although the objective function does not increase monotonically
at every iteration due to the optimization algorithm used, the in-situ adjoint procedure reaches final values comparable to those of the live finite-difference runs. Complementary optimization trajectories over the objective landscapes are shown
in Extended Data Fig.~\ref{fig:ed_opt3D}. The model-based trajectory in Fig.~\ref{fig4} serves as an offline theoretical
reference, rather than an additional hardware experiment. These results establish the functional feasibility of symmetry-driven in-situ adjoint gradients in closed-loop optimization on the original nonreciprocal hardware.

\section{Discussion}

Our results demonstrate that Lorentz reciprocity is
not a prerequisite for accessing adjoint dynamics
on the original physical system. Instead, a fixed
transpose mapping can establish the required
correspondence, provided that the transformed
excitations and readouts remain physically accessible
throughout parameter tuning. Importantly, this
symmetry need not impose global geometric constraints
on the entire system. For permutation symmetry,
the nontrivial transformation can be confined to
selected degrees of freedom, allowing arbitrarily
structured reciprocal subsystems connected through
permutation-invariant interfaces. Similarly, gauge
symmetry permits independent design and tuning of
reciprocal couplings within equal-weight sectors.
Schur reduction further reveals that the required
transpose symmetry may emerge only in the effective
dynamics of accessible degrees of freedom, even
without an apparent symmetry in the full system. These mappings do not restore
reciprocity: the experimental systems retain their
direction-dependent responses and operate under
unchanged nonreciprocal device.

The significance of these mappings extends beyond
establishing an algebraic relation between an
operator and its transpose. A port-level transpose
relation alone does not provide the local fields
required for gradients of internal parameters.
Our experiments bridge this gap by combining
measured port responses with calibrated local
relations to reconstruct the information needed
for gradient evaluation, without relying on
full-network forward and adjoint simulations.
The resulting in-situ adjoint gradients are
evaluated against independent experimental finite
differences and model-derived theoretical gradients,
providing complementary tests of gradient accuracy
and model consistency. Their use in live closed-loop
optimization further demonstrates that these
experimentally reconstructed gradients can guide
parameter updates on the original nonreciprocal
hardware, achieving performance comparable to
finite-difference control. Together, these results
connect symmetry-driven adjoint access to
functional physical optimization.

More broadly, the conditions for physical adjoint
computing extend beyond the existence of an
operator-level transpose relation to include the
accessibility of excitations, readouts and local
gradient information. Our results demonstrate
how distinct structured transpose mappings can be
constructed, experimentally accessed and used for
closed-loop optimization in nonreciprocal wave
systems. In particular, Schur reduction provides
a route to adjoint access when internal degrees
of freedom cannot be directly excited or measured,
while retaining their influence on the effective
dynamics.

The present implementation also has limitations.
The transformed adjoint responses are reconstructed
by digital superposition of sequential single-port
measurements rather than direct coherent excitation
of the transformed adjoint source. Gradient
evaluation requires calibrated local relations,
and the stabilized update directions need not
coincide with the exact derivatives of the
unmodified objective. Moreover, the demonstrated
systems contain only one or two independently
tunable parameters, and an advantage in total
measurement time over finite differences has not
yet been established. Extending this approach to
larger programmable systems will require fixed,
well-conditioned transpose mappings, scalable
excitation and readout, and reliable recovery of
local gradient information. Direct coherent
adjoint excitation and local-field measurements
offer possible routes towards more efficient
implementations. More broadly, these findings
identify experimentally accessible transpose
symmetry as a resource for gradient-based physical
computing while preserving nonreciprocal dynamics.

\section*{Methods}

\subsection*{Microwave platform and experimental calibration}

We assembled four nonreciprocal microwave networks (quantum graphs), each with four accessible ports, using coaxial transmission lines, T-junctions, ferrite-based nonreciprocal components and motorized phase shifters. The experimental setup comprised a four-port vector network analyzer (VNA; Ceyear 3657E), coaxial cables (HASCO SMA RG402), T-junctions (Pomona 72968), circulators (PE8403) and mechanical phase shifters (Narda P1501D) driven by step motors under Raspberry Pi control. The nominal coaxial inner-conductor radius and inner radius of the outer conductor are approximately 0.5~mm and 1.5~mm, respectively, and the relative dielectric permittivity is approximately 2.007. The phase shifters varied the effective propagation lengths of selected bonds while the nonreciprocal components remained unchanged. The phase to length conversion and discrete motor-control protocol are detailed in SI~\cite{symipac2026SI}. We define $S_{ij}$ as the outgoing complex-wave amplitude at port $i$ under unit-amplitude incidence at port $j$. At each parameter setting, the VNA sequentially excite the four accessible ports and measure the complete complex $4\times4$ scattering matrix $S$. Frequency sweeps covered 7 to 9~GHz, with 100 kHz frequency spacing and a 20 kHz intermediate-frequency bandwidth.

The mirror-permutation, rotation-permutation, latent symmetry and gauge symmetry networks are analysed at 8.0276, 7.9500, 8.3700 and 8.6400~GHz, respectively. The mirror and gauge networks are scanned over $96\times93$ points in $(\Delta L_{56},\Delta L_{78})$ and $96\times96$ points in $(\Delta L_{58},\Delta L_{67})$, respectively; the rotation and latent networks are each sampled at 96 values of $\Delta L_{78}$. Mechanical positions are converted into effective propagation length increments using the same calibration for the experimental measurements and model. The reference effective bond lengths used in the model are $(0.52427,0.58545)$~m for the mirror network, 0.50767~m for the rotational network, 0.48388~m for the latent network and $(0.58148,0.54574)$~m for the gauge network, in the parameter order specified above. The short port leads are 6.8~mm in the mirror, rotation and latent networks, and 7.5, 7.5, 6.2 and 6.2~mm in the gauge network. Further numerical specifications and transpose mapping checks for the four networks are provided in SI~\cite{symipac2026SI}.


\subsection*{Microwave graph model and port response}

The coaxial networks were modelled as linear quantum graphs describing electromagnetic wave propagation. The voltage along each coaxial bond obeys the one dimensional Helmholtz equation. Applying field continuity and the Kirchhoff--Neumann condition at junctions gives the dimensionless nodal matrix $M_0$ for the reciprocal propagation edges:
\begin{equation}
(M_0)_{uv} =
\begin{cases}
-\displaystyle\sum_{e\in\mathcal{E}_u}\cot(kL_e), & u=v,\\[5pt]
\displaystyle\sum_{e\in\mathcal{E}_{uv}}\csc(kL_e), & u\ne v.
\end{cases}
\tag{11}
\end{equation}
Here $k$ is the complex wavenumber, $L_e$ is the effective length of edge $e$, $\mathcal{E}_u$ denotes the set of edges incident on node $u$, and $\mathcal{E}_{uv}$ denotes the set of edges connecting nodes $u$ and $v$. Contributions from parallel edges are summed.

Nonreciprocal components were represented by their measured complex scattering matrices $S_v$, rather than by idealized circulator models. Under the adopted boundary-flux convention, each component contributes $i(I-S_v)(I+S_v)^{-1}$ to the nodal matrix, provided that $I+S_v$ is invertible. Using $Q_v$ to embed each component block into the global nodal space, the full nodal matrix and the open-system port relations are
\begin{equation}
\begin{aligned}
M &= M_0+i\sum_v Q_v(I-S_v)(I+S_v)^{-1}Q_v^T,\\
A &= M+iW^TW,\qquad
A\Phi=2iW^Ta,\qquad
S=-I+2iWA^{-1}W^T.
\end{aligned}
\tag{12}
\end{equation}
Here $Q_v$ maps the local component ports to the corresponding global nodal degrees of freedom, $W$ selects the four experimentally accessible ports, $\Phi$ is the nodal field vector, and $a$ denotes the incident port amplitudes. The matrix $A$ is the port-loaded wave operator, and $S$ is the full-network scattering matrix. The measured component responses were interpolated to the operating frequency before assembly. The same port ordering and normalization were used for modelling and measurement. When the model and measurement reference planes coincide, these relations give $W\Phi=(I+S)a$. Further details of the coaxial-edge and nonreciprocal-component constructions are provided in SI~\cite{symipac2026SI}.

The complex wavenumbers used for local-field reconstruction were $(-238.435+0.139i)$, $(-236.128+0.137i)$, $(-248.573+0.146i)$ and $(-256.592+0.148i)$~$\mathrm{m}^{-1}$ for the mirror, rotational, latent and gauge symmetry networks, respectively. These parameters describe the calibrated effective graph model, whereas the measured multiport responses provide the basis for experimental adjoint reconstruction.

\subsection*{Geometric transpose mappings}

We specify a geometric relabeling by a permutation
matrix $P$, defined by
$P\mathbf e_i=\mathbf e_{\pi(i)}$.
A geometric transpose symmetry of the nodal
operator requires
\begin{equation}
PM^{\mathrm T}(\mathbf p)P^{-1}=M(\mathbf p),\qquad
M_{\pi(i)\pi(j)}(\mathbf p)=M_{ji}(\mathbf p)
\quad\text{for all }i,j.
\label{eq:methods_geometric_isomorphism}
\end{equation}
The geometric constraint need not extend over
the entire system. To see this, consider a
partition into a subsystem containing the
nonreciprocal couplings and an attached
reciprocal subsystem:
\[
A=
\begin{pmatrix}
B&C\\
C^{\mathrm T}&R
\end{pmatrix},
\qquad
P=\operatorname{diag}(P_B,I).
\]
The transpose relation is then equivalent to
\[
P_BB^{\mathrm T}P_B^{-1}=B,\qquad
P_BC=C,\qquad
R^{\mathrm T}=R.
\]
Thus, the reciprocal subsystem $R$ may have
arbitrary internal structure and parameter
dependence, provided the interface couplings
remain invariant under $P_B$. In particular,
when $P_B$ exchanges two terminals of a
nonreciprocal component while leaving a third
terminal fixed, an arbitrary reciprocal
subsystem can be attached through that
invariant terminal without breaking the
transpose mapping.  

The reversed system must be isomorphic
to the original one, including reciprocal and
directed couplings and diagonal self-actions.
For the same permutation to intertwine the
port-loaded operator $A$, the loading must also
obey $PW^{\mathrm T}WP^{-1}=W^{\mathrm T}W$
and the relevant source and readout ports must
transform consistently. The permutation is held
fixed across the admissible tuning family, rather
than recomputed separately at each parameter setting.

Mirror and rotational geometric constructions
both induce the retained-port exchange $P_{12}$,
which swaps ports 1 and 2 and leaves ports 3
and 4 unchanged. Equivalently, $P$ leaves
$(M+M^{\mathrm T})/2$ invariant and reverses
the sign of $(M-M^{\mathrm T})/2$.
A constructive proof for general weighted
permutations is provided in SI~\cite{symipac2026SI}.

\subsection*{Complex gauge reweighting and component calibration}

For a diagonal transformation
$\Lambda=\operatorname{diag}(d_1,\ldots,d_N)$
with all $d_i\neq0$, the condition
$\Lambda A^{\mathrm T}\Lambda^{-1}=A$
requires equal nonzero support in both coupling
directions and
\begin{equation}
\frac{A_{ij}(\mathbf p)}{A_{ji}(\mathbf p)}
=\frac{d_i}{d_j}
\quad (A_{ij}A_{ji}\neq0).
\label{eq:methods_local_gauge_ratio}
\end{equation}
Reciprocal connections require equal endpoint
weights, whereas the complex ratios of nonreciprocal
couplings determine relative weights between node
groups. A globally consistent assignment additionally
requires
$\prod_{(i,j)\in C}A_{ij}/A_{ji}=1$
along each oriented independent graph cycle $C$.
Weights can then be propagated from one reference
node along a spanning tree; the associated
constructive proof is given in SI~\cite{symipac2026SI}. The same ratios, and hence the same
$\Lambda$, must remain valid throughout the
prescribed parameter changes.

For the complex-weighted prototype, the nontrivial
weight was obtained from measured scattering
parameters of the directional component, rather
than fitted to the full-network scattering matrix.
With $S_v^{(2)}$ denoting its equivalent two-port
scattering representation in the calibrated
graph model, we used
\begin{equation}
K_v=i\bigl(I-S_v^{(2)}\bigr)
\bigl(I+S_v^{(2)}\bigr)^{-1},\qquad
q=\frac{(K_v)_{u\ell}}{(K_v)_{\ell u}}
\simeq0.8720+0.0989i.
\label{eq:methods_calibrated_q}
\end{equation}
Here $u$ and $\ell$ label node groups assigned
weights $q$ and $1$, respectively, under the
chosen coupling orientation. Since $|q|\neq1$,
this is both an amplitude and a phase reweighting.
The port-space map is
$D_{\mathrm p}=\operatorname{diag}(q,1,1,q)$.
The independently tuned reciprocal branches
$L_{58}$ and $L_{67}$ connect nodes satisfying
$d_5=d_8$ and $d_6=d_7$, respectively;
changing their lengths therefore preserves
the weight relations provided the
directional-component response is unchanged.


\subsection*{Schur reduction and latent transpose symmetry}

Only the external ports are independently accessible
for excitation and readout. We therefore partition the
port-loaded operator into retained states $r$ and
inaccessible internal states $i$, with no independent
internal source. If $A_{ii}$ is invertible, exact
elimination gives
\begin{equation}
A_{\mathrm{eff}}
=A_{rr}-A_{ri}A_{ii}^{-1}A_{ir},\qquad
A_{\mathrm{eff}}\boldsymbol{\Phi}_r=\mathbf b_r.
\label{eq:methods_schur_operator}
\end{equation}
The self-energy term retains all multiple-scattering
feedback from the eliminated states; the physical
internal network is not removed. For an objective
and adjoint source confined to the retained ports,
the reduced adjoint equation is
$A_{\mathrm{eff}}^{\mathrm T}
\boldsymbol{\lambda}_r=\mathbf c_r$.
Its response can be recovered through a fixed
port-space transformation if
\begin{equation}
D_{\mathrm{eff}}A_{\mathrm{eff}}^{\mathrm T}(\mathbf p)
D_{\mathrm{eff}}^{-1}=A_{\mathrm{eff}}(\mathbf p)
\quad\text{for all }\mathbf p\in\mathcal P.
\label{eq:methods_latent_transpose}
\end{equation}

To establish when a full-space transpose mapping
is inherited by the reduced system, suppose a fixed
$D$ satisfies $A(\mathbf p)D=D A^{\mathrm T}(\mathbf p)$
throughout the admissible tuning family, and partition
$D$ in the same retained/internal basis:
\begin{equation}
D=\begin{pmatrix}
D_{rr}&D_{ri}\\
D_{ir}&D_{ii}
\end{pmatrix}.
\label{eq:methods_D_block_partition}
\end{equation}
If the transformation preserves the partition,
$D_{ri}=D_{ir}=0$, so that
$D=\operatorname{diag}(D_r,D_i)$.
The corresponding block identities imply
$A_{rr}D_r=D_rA_{rr}^{\mathrm T}$,
$A_{ir}D_r=D_iA_{ri}^{\mathrm T}$,
$A_{ri}D_i=D_rA_{ir}^{\mathrm T}$, and
$A_{ii}^{-1}D_i=D_iA_{ii}^{-\mathrm T}$.
Substituting these into the Schur complement gives
\begin{equation}
\begin{aligned}
A_{\mathrm{eff}}D_r
&=(A_{rr}-A_{ri}A_{ii}^{-1}A_{ir})D_r\\
&=D_r\bigl(A_{rr}^{\mathrm T}
 -A_{ir}^{\mathrm T}A_{ii}^{-\mathrm T}
 A_{ri}^{\mathrm T}\bigr)\\
&=D_rA_{\mathrm{eff}}^{\mathrm T}.
\end{aligned}
\label{eq:methods_schur_inheritance}
\end{equation}
Hence $D_{\mathrm{eff}}=D_r$ (denoted
$D_{\mathrm p}$ when $r$ contains the four ports)
provides the same transpose similarity after reduction.

This conclusion does not extend directly to a general
$D$ that mixes retained and internal states.
Let $\mathbf c=(\mathbf c_r,0)^{\mathrm T}$,
$\mathbf y=D\boldsymbol{\lambda}$, and
$A^{\mathrm T}\boldsymbol{\lambda}=\mathbf c$.
The internal adjoint equation yields
\begin{equation}
\begin{aligned}
\boldsymbol{\lambda}_i
&=-A_{ii}^{-\mathrm T}A_{ri}^{\mathrm T}
\boldsymbol{\lambda}_r,\\
\mathbf y_r
&=\bigl(D_{rr}-D_{ri}A_{ii}^{-\mathrm T}
A_{ri}^{\mathrm T}\bigr)\boldsymbol{\lambda}_r
\equiv D_{\mathrm{fld}}\boldsymbol{\lambda}_r.
\end{aligned}
\label{eq:methods_schur_field_map}
\end{equation}
Meanwhile, the transformed equation
$A\mathbf y=D\mathbf c$ has an induced internal
source $D_{ir}\mathbf c_r$. Eliminating its internal
field gives
\begin{equation}
\begin{aligned}
\mathbf y_i
&=A_{ii}^{-1}(D_{ir}\mathbf c_r
 -A_{ir}\mathbf y_r),\\
A_{\mathrm{eff}}\mathbf y_r
&=\bigl(D_{rr}-A_{ri}A_{ii}^{-1}D_{ir}\bigr)
\mathbf c_r
\equiv D_{\mathrm{src}}\mathbf c_r.
\end{aligned}
\label{eq:methods_schur_source_map}
\end{equation}
Because $\mathbf c_r=A_{\mathrm{eff}}^{\mathrm T}
\boldsymbol{\lambda}_r$, these relations imply
\begin{equation}
A_{\mathrm{eff}}D_{\mathrm{fld}}
=D_{\mathrm{src}}A_{\mathrm{eff}}^{\mathrm T}.
\label{eq:methods_schur_two_sided}
\end{equation}
Thus the effective excitation is
$D_{\mathrm{src}}\mathbf c_r$, whereas recovering
the adjoint field requires
$\boldsymbol{\lambda}_r
=D_{\mathrm{fld}}^{-1}\mathbf y_r$, provided
$D_{\mathrm{fld}}$ is invertible and both operations
are experimentally accessible. In general,
$D_{\mathrm{src}}\neq D_{\mathrm{fld}}$; they can
also depend on $\mathbf p$ even when $D$ is fixed.
When $D_{ri}=D_{ir}=0$, both reduce to $D_r$,
recovering Eq.~\eqref{eq:methods_schur_inheritance}.
A boundary-equivalent excitation reproduces the
retained response, but need not reproduce the internal
field of the transformed full-space problem.

A latent transpose symmetry is a separate possibility:
a simple fixed $D_{\mathrm{eff}}$ can satisfy
Eq.~\eqref{eq:methods_latent_transpose} without
being inherited from a physically accessible
full-space transformation. This distinction is
related to latent symmetries revealed by isospectral
reduction \cite{kempton2020isospectral,rontgen2021latent,
rontgen2023equireflectionality}.
The latent-symmetry graph contains 22 modelled nodal
degrees of freedom, including four experimentally
accessible ports. Its effective mapping
$D_{\mathrm{eff}}=P_{12}$ exchanges ports 1 and 2,
while leaving ports 3 and 4 unchanged, throughout
the measured $L_{78}$ scan at 8.3700 GHz.
The mapping is checked against every matrix element
of the reduced operator, not solely the two exchanged
diagonal terms. Unlike the mirror implementation,
the candidate permutation is identified after internal
elimination.

Schur reduction provides the accessible adjoint response
but does not automatically supply internal fields.
If local sensitivities require eliminated states,
their forward fields follow from the calibrated blocks
as $\boldsymbol{\Phi}_{i}=-A_{ii}^{-1}A_{ir}
\boldsymbol{\Phi}_{r}$, and their adjoint fields
follow from Eq.~\eqref{eq:methods_schur_field_map}.
Equivalently, the corresponding parameter derivative
of $A_{\mathrm{eff}}$ must include the full derivative
of the self-energy. Further details are given in
SI~\cite{symipac2026SI}.

\subsection*{Port-level adjoint reconstruction and transpose test}

For full-space maps compatible with the source
and measurement embedding,
$DW^{\mathrm T}=W^{\mathrm T}D_{\mathrm p}$
and $WD=D_{\mathrm p}W$; for latent symmetries,
the same input--output relation follows directly
from the reduced operator. In both cases the
measurable signature is
$SD_{\mathrm p}=D_{\mathrm p}S^{\mathrm T}$.
We evaluate it using the complete measured
four-port response and the normalized residual
\begin{equation}
r_D(\mathbf p)=
\frac{\bigl\|D_{\mathrm p}S^{\mathrm T}(\mathbf p)
D_{\mathrm p}^{-1}-S(\mathbf p)\bigr\|_{\mathrm F}}
{\bigl\|S(\mathbf p)\bigr\|_{\mathrm F}}.
\label{eq:rD}
\end{equation}
Here $D_{\mathrm p}=P_{12}$ for the geometric
and latent devices and
$D_{\mathrm p}=\operatorname{diag}(q,1,1,q)$
for the complex-weighted device. The transformations
were fixed within each sampled parameter family.
The boundary residual tests the measured transpose
relation, not the accuracy of the internal-field
reconstruction.

For the directional-power objective
$g_\sigma=\sigma(|\varphi_2^{(1)}|^2
-|\varphi_1^{(2)}|^2)$, where $\sigma=-1$
for the mirror network and $+1$ otherwise,
$\varphi_m^{(n)}$ denotes the field response at
port $m$ under unit incidence at port $n$.
In our normalization, $\varphi_2^{(1)}=S_{21}$
and $\varphi_1^{(2)}=S_{12}$.
The adjoint equation uses
$A^{\mathrm T}\boldsymbol{\lambda}=\mathbf c$,
with ordinary transpose rather than conjugate
transpose. The two forward excitations produce
the objective-dependent sources
\begin{equation}
\mathbf c^{(1)}
=-\sigma\varphi_2^{(1)*}\mathbf e_2,\qquad
\mathbf c^{(2)}
=\sigma\varphi_1^{(2)*}\mathbf e_1.
\label{eq:methods_adjoint_sources}
\end{equation}
For either source, the effective transformed
problem is
$A_{\mathrm{eff}}\boldsymbol{\lambda}'_r
=D_{\mathrm p}\mathbf c_r$ and
$\boldsymbol{\lambda}_r
=D_{\mathrm p}^{-1}\boldsymbol{\lambda}'_r$.

When the model and measured port reference planes
are aligned, Eq.~12 gives the response
to unit incidence at port $j$ as
$\boldsymbol{\Phi}_r^{(j)}=(I+S)\mathbf e_j$.
Consequently, the field associated with the
transformed source can be synthesized directly
from the measured basis:
\begin{equation}
\boldsymbol{\lambda}'_r
=\sum_{j=1}^{4}
\frac{(D_{\mathrm p}\mathbf c_r)_j}{2i}
\boldsymbol{\Phi}_r^{(j)}
=\frac{I+S}{2i}D_{\mathrm p}\mathbf c_r.
\label{eq:methods_basis_synthesis}
\end{equation}
We use sequential VNA port excitations followed
by numerical complex superposition; a simultaneous
coherent multiport launch is not performed.
When short leads separate reference planes,
their calibrated response must be included
before applying this port-space reconstruction,
as detailed in SI~\cite{symipac2026SI}.
The measured responses determine the retained-port
propagation, whereas calibrated local graph relations
are needed for the internal fields and matrix
derivatives entering local sensitivities.

\subsection*{Local sensitivities, stabilization and validation}

For fixed incident excitations and port coupling,
the raw adjoint sensitivity of the directional-power
objective to a bond length is
\begin{equation}
G_e=2\operatorname{Re}\sum_{s=1}^{2}
(\boldsymbol{\lambda}^{(s)})^{\mathrm T}
\frac{\partial M}{\partial L_e}
\boldsymbol{\Phi}^{(s)}.
\label{eq:rawG}
\end{equation}
This contraction uses the forward and adjoint fields
at the endpoints of the tuned edge. The experimentally
reconstructed local fields are combined with the
calibrated operator derivative; the procedure does
not use purely numerical forward and adjoint solves
in place of the measured port responses.

The nodal representation can become poorly conditioned
near small $|\sin(kL_e)|$, amplifying errors in
local-field reconstruction. A predetermined,
parameter-dependent stabilization is therefore
applied before the sensitivity is supplied to
the controller:
\begin{equation}
\widetilde G_e=w_eG_e,\qquad
w_e=\frac{|\sin(kL_e)|^2}
{|\sin(kL_e)|^2+\epsilon_e^2}.
\label{eq:weight}
\end{equation}
The fixed regularization parameters $\epsilon_e$
are $(0.22,0.08)$ for $(L_{56},L_{78})$ in the
mirror network, $0.10$ for $L_{78}$ in each of
the rotational and latent networks, and
$(0.25,0.25)$ for $(L_{58},L_{67})$ in the
complex-weighted network. The weight $w_e$,
unlike $\epsilon_e$, changes with the instantaneous
design state. This is a sensitivity preconditioning
operation, not smoothing of plotted measurements;
in general $\widetilde G_e$ is not the exact
derivative of the unmodified objective.

We compared $\widetilde G_e$ with gradients
independently estimated from measured objective
values. For the scanned parameter grids, the
central finite-difference reference at interior
points is
\begin{equation}
G_k^{\mathrm{FD}}(\mathbf p)=
\frac{g(\mathbf p+h_k\mathbf e_k)
-g(\mathbf p-h_k\mathbf e_k)}{2h_k},
\label{eq:methods_fd}
\end{equation}
with nominal displacement $h_k\simeq0.2961$ mm
per side and calibrated length coordinates used
in analysis. For the separate live
finite-difference optimization, the mirror and
rotational networks use one motor step on each
side of the working point; the complex-weighted
and latent networks use a forward step.

We quantify agreement by Pearson's correlation
coefficient $R$ and by the range-normalized
root-mean-square error
\begin{equation}
\mathrm{NRMSE}
=100\%\,
\frac{\sqrt{N^{-1}\sum_{j=1}^{N}
(\widetilde G_j-G_j^{\mathrm{FD}})^2}}
{\max_jG_j^{\mathrm{FD}}-\min_jG_j^{\mathrm{FD}}}.
\label{eq:methods_nrmse}
\end{equation}
Two-dimensional statistics include all supplied
valid paired scan points. These grid points are
not independent experimental repeats; correlations
and NRMSEs are therefore descriptive comparisons,
not confidence intervals.


\subsection*{Closed-loop optimization and measurement cost}

Hardware optimization uses the NLopt implementation
of the Method of Moving Asymptotes (LD\_MMA)
\cite{svanberg1987,johnson2021nlopt}, with a relative objective
tolerance of $10^{-3}$ and the recorded evaluation
budgets. For each live-adjoint update, the motorized
phase shifters move to an attainable setting, the
VNA acquires the complex port response, and the
transformed adjoint and stabilized update direction
are reconstructed before requesting the next setting.
In live finite-difference optimization, the controller
instead perturbs the tunable lengths and obtains the
update direction from newly measured objective values.

The comparison trajectories use model-database, live measured-basis adjoint and
live finite-difference updates. Database queries
use the nearest recorded calibrated grid coordinate;
live trajectories retune and remeasure the hardware.
Figure~\ref{fig4} displays the offline model-database
reference alongside the two live trajectories;
the measured-database endpoints are also reported
in Extended Data Table~\ref{tab:ed_optimization}.
Results report the recorded final objective values
rather than replacing them with the maximum
attained earlier in a trajectory.

At a fixed number of excitation and objective states,
the number of mathematical forward/adjoint solves
does not increase with the number of tunable
parameters. The present experimental implementation,
however, measures a complete four-port basis
sequentially and synthesizes the transformed response
digitally. Its VNA sweep count and wall-clock
acquisition time therefore depend on the measurement
architecture and cannot be inferred from the
mathematical adjoint solve count. No experimentally
measured speedup relative to finite differences
is claimed. The regularization, finite-difference protocols
and optimization bookkeeping are further documented
in SI~\cite{symipac2026SI}.


\bibliographystyle{naturemag}
\bibliography{References}


\clearpage
\begingroup
\setcounter{figure}{0}
\setcounter{table}{0}
\renewcommand{\thefigure}{\arabic{figure}}
\renewcommand{\thetable}{\arabic{table}}
\renewcommand{\theHfigure}{EDfig.\arabic{figure}}
\renewcommand{\theHtable}{EDtable.\arabic{table}}
\renewcommand{\figurename}{Extended Data Fig.}
\renewcommand{\tablename}{Extended Data Table}
\captionsetup[figure]{font=small,labelfont=bf,labelsep=colon,
  justification=raggedright,singlelinecheck=false}
\captionsetup[table]{font=small,labelfont=bf,labelsep=colon,
  justification=raggedright,singlelinecheck=false,position=top}

\begin{figure}[H]
\centering
\includegraphics[width=0.98\textwidth]{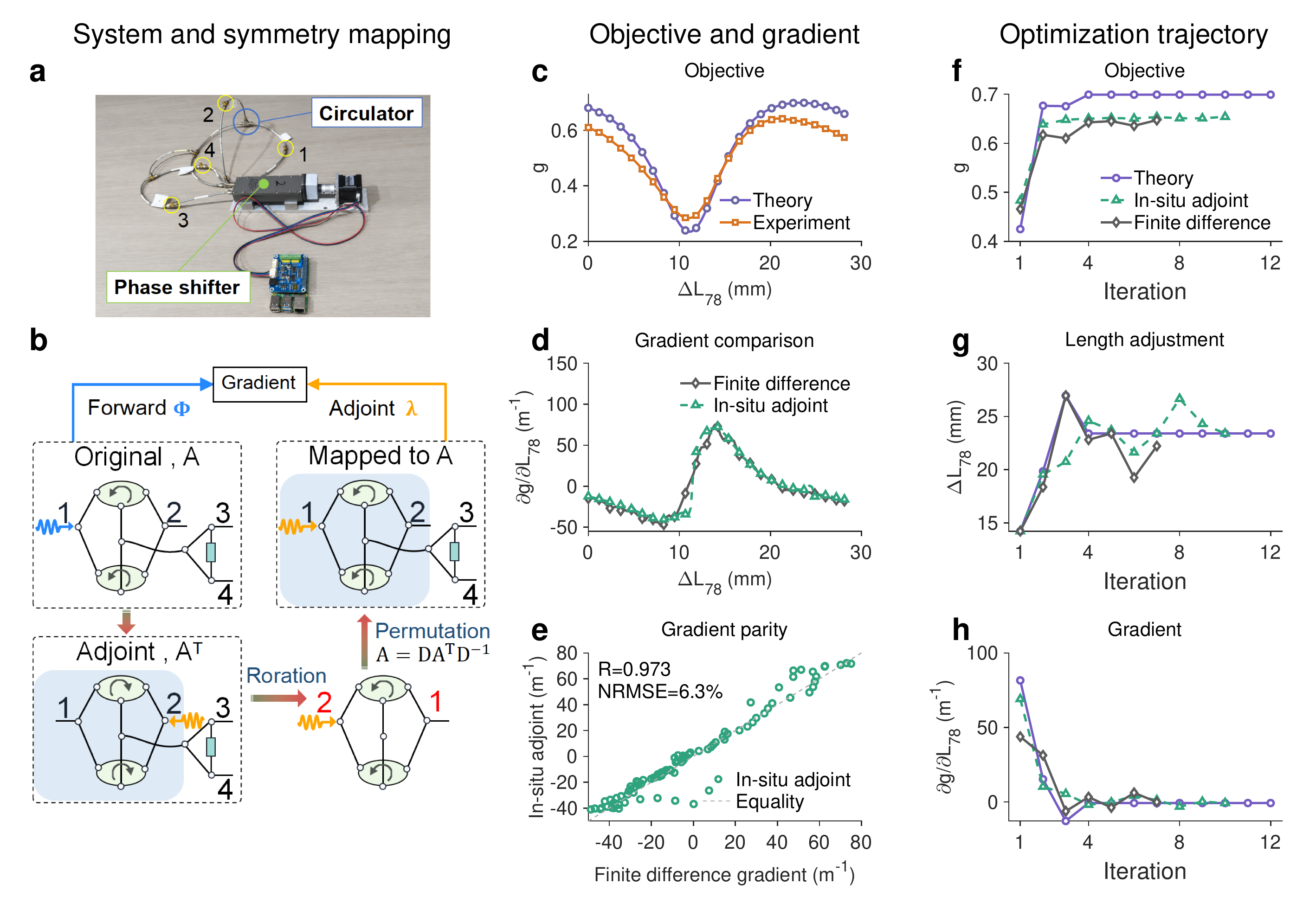}
\caption{
\textbf{Rotational transpose symmetry and in-situ adjoint optimization in a nonreciprocal microwave network.}
\textbf{a}, Photograph of the four-port microwave network. Yellow circles and numbers identify the accessible ports; the circulator and tunable phase shifter are indicated.
\textbf{b}, Schematic of the rotational transpose mapping. A $180^\circ$ rotation induces the port permutation $D=P_{12}$, which maps the adjoint operator $A^{T}$ onto the original operator $A$ according to $A=DA^{T}D^{-1}$, without reversing the nonreciprocal components.
\textbf{c}, Directional-transmission objective $g=|S_{21}|^2-|S_{12}|^2$ as a function of the bond length adjustment $\Delta L_{78}$, comparing measurements with the calibrated quantum-graph model.
\textbf{d,} Stabilized in-situ adjoint gradients compared with independently measured finite-difference gradients across the $\Delta L_{78}$ scan.
\textbf{e,} Pointwise comparison between stabilized in-situ adjoint gradients and finite-difference estimates. The dashed line indicates equality; $R$ denotes the Pearson correlation coefficient and NRMSE the range-normalized root-mean-square error.
\textbf{f,} Evolution of the objective during optimization using in-situ adjoint gradients and live finite differences. The model-based trajectory is shown as an offline reference.
\textbf{g,} Evolution of the branch-length adjustment $\Delta L_{78}$ during the corresponding optimization trajectories.
\textbf{h,} Gradient estimates along the optimization trajectories. In c, f and h, the theory curve represents an offline model-based reference, whereas the in-situ adjoint and finite-difference curves represent live hardware optimization. The in-situ adjoint gradients are stabilized as described in Methods.
}
\label{fig:ed_rotation}
\end{figure}
\clearpage

\begin{figure}[H]
\centering
\includegraphics[width=\textwidth]{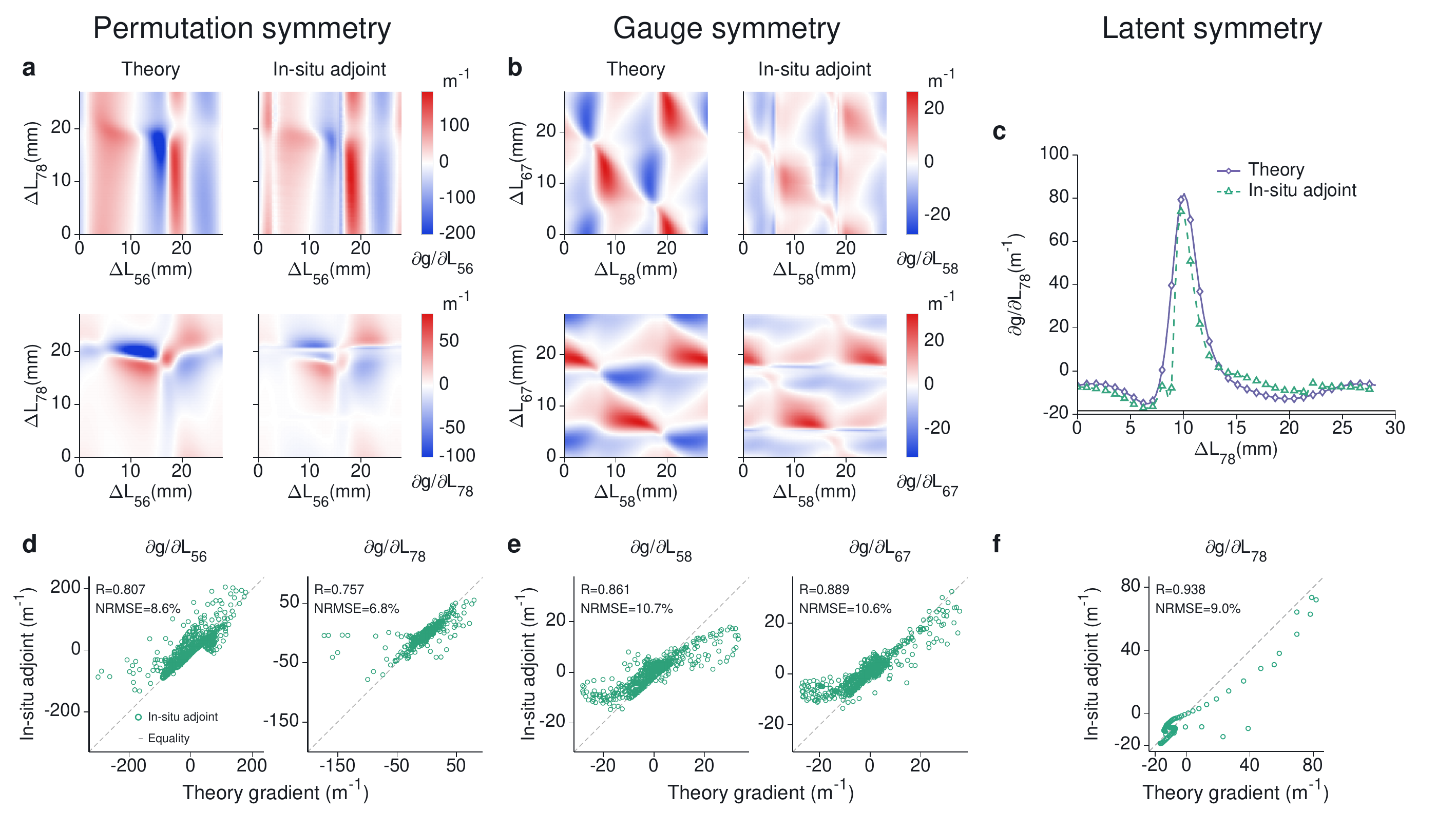}
\caption{\textbf{Comparison of theoretical and in-situ adjoint gradients.}
\textbf{a,b,} Distributions of theoretical and in-situ adjoint gradients for the permutation symmetry (\textbf{a}) and gauge symmetry (\textbf{b}) networks. The upper and lower rows show gradients with respect to the two tunable bond lengths in each network. \textbf{c,} Theoretical and in-situ adjoint gradients as a function of $\Delta L_{78}$ for the latent symmetry network. \textbf{d--f,} Pointwise comparisons between theoretical and in-situ adjoint gradients for the permutation symmetry (\textbf{d}), gauge symmetry (\textbf{e}) and latent symmetry (\textbf{f}) networks. Dashed lines indicate equality. $R$ denotes the Pearson correlation coefficient, and NRMSE the range-normalized root-mean-square error. The in-situ adjoint gradients include the stabilization described in Methods.}

\label{fig:ed_theory}
\end{figure}
\clearpage

\begin{figure}[H]
\centering
\includegraphics[width=\textwidth]{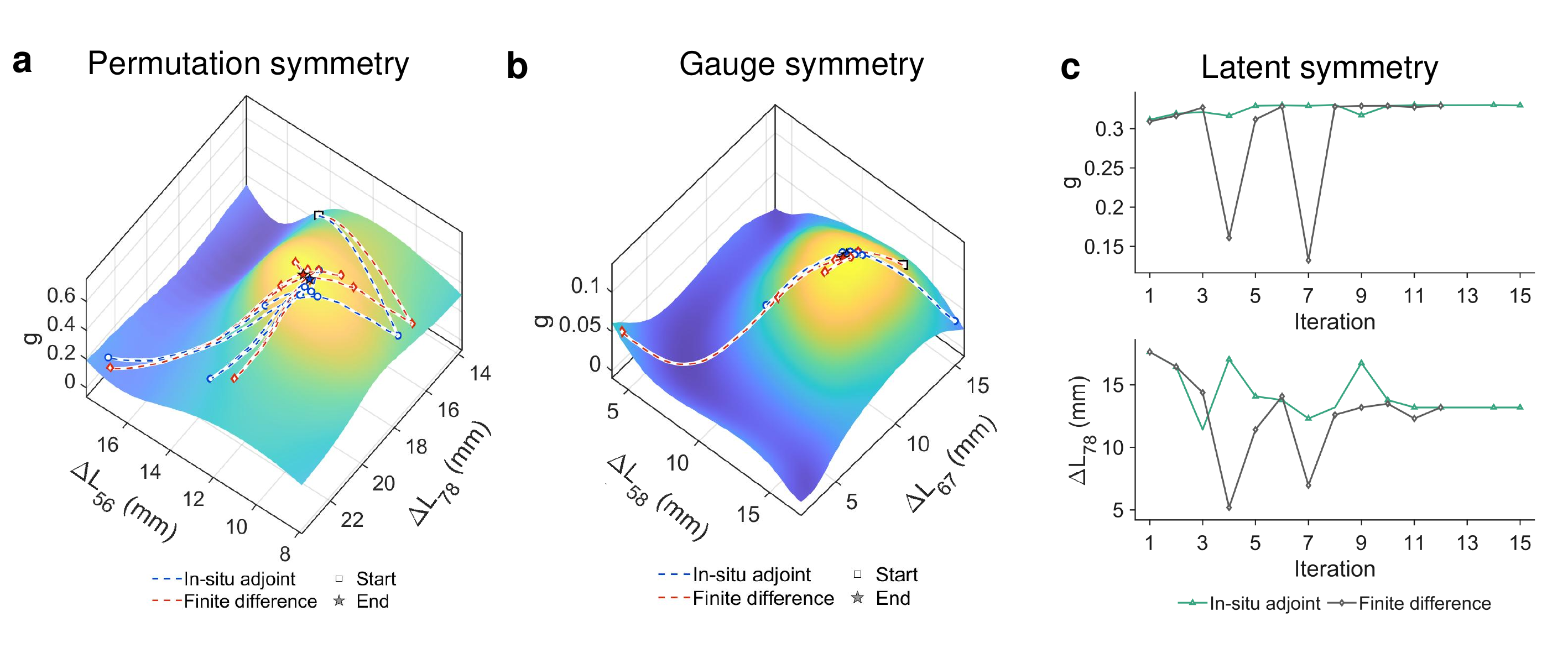}

\caption{\textbf{Closed-loop optimization of directional wave transport in nonreciprocal microwave networks.} \textbf{a,b,} Objective landscapes and experimental optimization trajectories for the permutation (mirror symmetric) (\textbf{a}) and complex gauge (\textbf{b}) networks. Blue and red dashed lines indicate live adjoint-based and finite-difference optimization, respectively. Open squares and stars mark the initial and final parameter settings. \textbf{c,} Evolution of the objective $g$ (top) and tunable branch length $\Delta L_{78}$ (bottom) during closed-loop optimization of the hidden-symmetry network. Green triangles and grey diamonds represent live adjoint-based and finite-difference updates, respectively. All experiments are performed on the original nonreciprocal hardware.}

\label{fig:ed_opt3D}
\end{figure}
\clearpage
\begin{table}[H]
\centering
\caption{\textbf{Measured transformed-transpose residuals.}}

\small
\setlength{\tabcolsep}{9pt}
\renewcommand{\arraystretch}{1.18}
\begin{tabular}{@{}lrrrrrr@{}}
\toprule
Network & \(N\) & Mean & Median & SD & Minimum & Maximum \\
\midrule
Mirror           & 8,928 & 0.06080 & 0.06137 & 0.01933 & 0.01509 & 0.10793 \\
Hidden           &    96 & 0.06774 & 0.06845 & 0.00801 & 0.04803 & 0.07778 \\
Complex-weighted & 9,216 & 0.01150 & 0.01107 & 0.00285 & 0.00493 & 0.02336 \\
Rotation         &    96 & 0.10890 & 0.12090 & 0.02333 & 0.05263 & 0.15768 \\
\bottomrule
\label{tab:ed_residuals}
\end{tabular}
\par\vspace{6pt}
\begin{minipage}{0.96\linewidth}
\footnotesize
For each measured four-port scattering matrix \(S\), the dimensionless residual is
\(r_D=\|D_{\mathrm p}S^{\mathrm T}D_{\mathrm p}^{-1}-S\|_{\mathrm F}/\|S\|_{\mathrm F}\).
The port transformation is \(P_{12}\) for the mirror, hidden and rotational networks
and \(\operatorname{diag}(q,1,1,q)\) for the complex-weighted network, with
\(q\simeq0.8720+0.0989i\). \(N\) is the number of measured parameter settings;
SD is the standard deviation across these settings, \emph{not} uncertainty across
independent experimental replicates. The same port map is used throughout each scan.
\end{minipage}
\end{table}
\clearpage

\begin{table}[H]
\centering
\caption{\textbf{Agreement between  in-situ adjoint gradients and finite differences.}}
\label{tab:ed_gradients}\label{tab:gradmetrics}
\small
\setlength{\tabcolsep}{15pt}
\renewcommand{\arraystretch}{1.18}
\begin{tabular}{@{}llrr@{}}
\toprule
Network & Tunable length & Pearson \(R\) & NRMSE (\%) \\
\midrule
Mirror           & \(L_{56}\) & 0.8950 & 7.6 \\
                 & \(L_{78}\) & 0.7760 & 6.0 \\
Hidden           & \(L_{78}\) & 0.87630 & 10.851 \\
Gauge            & \(L_{58}\) & 0.86282 & 9.776 \\
                 & \(L_{67}\) & 0.86014 & 11.450 \\
Rotation         & \(L_{78}\) & 0.97330 & 6.295 \\
\bottomrule
\end{tabular}
\par\vspace{6pt}
\begin{minipage}{0.94\linewidth}
\footnotesize
Pearson \(R\) and the range-normalized root-mean-square error (NRMSE) compare the
\emph{stabilized measured-adjoint update direction} \(\widetilde G\) with the
independently measured finite-difference sensitivity \(G^{\mathrm{FD}}\).
NRMSE is \(100\%\times\operatorname{RMSE}(\widetilde G-G^{\mathrm{FD}})\)
divided by the range of \(G^{\mathrm{FD}}\) within the corresponding scan.
Stabilized updates need not equal the unmodified analytic gradient.
The values summarize scan-point agreement and are not confidence intervals from
independent repeats.
\end{minipage}
\end{table}
\clearpage

\begin{table}[H]
\centering
\caption{\textbf{Objective values from database and live optimization trajectories.}}
\label{tab:ed_optimization}\label{tab:endpoints}\label{tab:live}
\footnotesize
\setlength{\tabcolsep}{6.3pt}
\renewcommand{\arraystretch}{1.2}
\begin{tabular}{@{}lrrrrrr@{}}
\toprule
& \multicolumn{2}{c}{Database endpoint} & \multicolumn{3}{c}{Live adjoint trajectory} & Live FD \\
\cmidrule(lr){2-3}\cmidrule(lr){4-6}\cmidrule(l){7-7}
Network & Model & Measured & Initial & Final & Maximum & Final \\
\midrule
Mirror           & 0.749949 & 0.646046 & 0.278235 & 0.617420 & 0.617420 & 0.613154 \\
Hidden           & 0.327412 & 0.327156 & 0.311476 & 0.329824 & 0.330438 & 0.329370 \\
Complex-weighted & 0.151339 & 0.116922 & 0.086882 & 0.115096 & 0.116818 & 0.116316 \\
Rotation         & 0.699063 & 0.654252 & 0.483284 & 0.654008 & 0.654008 & 0.647122 \\
\bottomrule
\end{tabular}
\par\vspace{6pt}
\begin{minipage}{0.96\linewidth}
\footnotesize
Values are dimensionless directional-transport objectives,
\(g=\sigma(|S_{21}|^2-|S_{12}|^2)\), with \(\sigma=-1\) for the mirror network and
\(\sigma=+1\) otherwise. Database endpoints are the terminal values of the
model- and measurement-database optimization traces; they are not global database
maxima. The live-adjoint initial, final and maximum values refer to the same
recorded trajectory. In particular, the hidden-network endpoint (0.329824) differs
from its maximum (0.330438). Live finite differences (FD) report their trajectory
endpoints. These are selected single-run records, not replicate means; the database
and live paths do not have identical acquisition costs.
\end{minipage}
\end{table}
\clearpage
\endgroup

\end{document}